# 3D Printed Hybrid Refractive/Diffractive Achromat and Apochromat for the Visible Wavelength Range

MICHAEL SCHMID[1*], FLORIAN STERL[1], SIMON THIELE[2], ALOIS HERKOMMER[2], AND HARALD GIESSEN[1]

[1]4th Physics Institute and Research Center SCoPE, University of Stuttgart, Pfaffenwaldring 57, 70569 Stuttgart, Germany.
[2]Institute of Applied Optics (ITO) and Research Center SCoPE, University of Stuttgart, Pfaffenwaldring 9, 70569 Stuttgart, Germany.
*Corresponding author: m.schmid@pi4.uni-stuttgart.de

**3D direct laser writing is a powerful technology to create nano- and microscopic optical devices. While the design freedom of this technology offers the possibility to reduce different monochromatic aberrations, reducing chromatic aberrations is often neglected. In this paper we successfully demonstrate the combination of refractive and diffractive surfaces to create a refractive/diffractive *achromat* and also show the first refractive/diffractive *apochromat* by using DOEs and simultaneously combining two different photoresists, namely IP-S and IP-n162. These combinations drastically reduce chromatic aberrations in 3D-printed micro-optics for the visible wavelength range. The optical properties as well as the substantial reduction of chromatic aberrations are characterized and we outline the benefits of three-dimensional direct laser written achromats and apochromats for micro-optics.** © 2021 Optical Society of America

http://doi.org/10.1364/OL.423196

3D direct laser writing is an important technology for the manufacturing of nano- and micro-optical elements [1,2] offering the possibility to create complex micro-optics with different functionality. First promising results towards applications, such as optical communication [3–5], beam shaping [6–8], microscopy [9], particle trapping [10,11], quantum coupling [12], sensing [13,14], imaging [15,16], or endoscopy [17,18] have been reported.

To improve the optical performance of these devices, different aberrations need to be reduced. Monochromatic aberrations such as defocus, spherical aberration, coma, astigmatism, field curvature or image distortion can easily be reduced with 3D direct laser writing due to the large design freedom, thus enabling the use of aspheric surfaces or additional optical surfaces to realize multi-lens objectives [19,20]. Chromatic aberrations (longitudinal chromatic aberration (LCA) and transverse chromatic aberration (TCA)) cannot be corrected with these methods. In principle, two approaches can be used to reduce chromatic aberrations, namely the combination of photoresists with different dispersion, or the use of diffractive optical surfaces on top of a refractive surfaces [21,22]. A reduction of the chromatic aberrations in 3D direct laser written micro-optics has already been demonstrated by combining two photoresists with different dispersion creating a multi-component Fraunhofer achromat [23].

In the following we demonstrate the achromatic function of a refractive surface combined with a diffractive surface, as well as the combination of both approaches, which result in the first 3D direct laser written refractive/diffractive multi-material apochromat, reducing the chromatic aberrations even further. While achromats that eliminate chromatic aberration for two wavelengths still exhibit a secondary chromatic spectrum, *apochromats* eliminate chromatic aberration completely for three different wavelengths.

Achromaticity for two wavelengths is achieved when the Fraunhofer condition is fulfilled: $F_1/v_1 + F_2/v_2 = 0$, with $F_i$ as the optical power of the lenses and $v_i$ as the Abbe numbers. To achieve apochromaticity the equation is expanded by a third term, reading $F_1/v_1 + F_2/v_2 + F_3/v_3 = 0$. The Abbe number $v = (n_d - 1)/(n_F - n_C)$ is an approximate measure for the strength of dispersion. The Abbe number for diffractive optical elements is independent of the lens shape and only depends on the wavelength band of the used light $v = \lambda_c/(\lambda_s - \lambda_l)$ with $\lambda_c$ the central wavelength, $\lambda_s$ the shorter wavelength and $\lambda_l$ the longer wavelength. It is always negative and comparably strong, e.g. $v = \lambda_c/(\lambda_s - \lambda_l) = 587.6/(486.1 - 656.3) = -3.45$ for the d-, F-, and C-line [22].

**Table 1. Refractive indices of the photoresists IP-S and IP-n162 in the visible range (500 nm - 900 nm)** [24]

| n | $n_{500\,nm}$ | $n_{600\,nm}$ | $n_{700\,nm}$ | $n_{800\,nm}$ | $n_{900\,nm}$ |
|---|---|---|---|---|---|
| IP-S | 1.5173 | 1.5104 | 1.5069 | 1.5047 | 1.5030 |
| IP-n162 | 1.6382 | 1.6221 | 1.6131 | 1.6079 | 1.6042 |

To obtain the required detailed knowledge of the dispersion in the visible light range we measured the refractive indices using an automated Pulfrich refractometer and display them in Table 1 [24]. The photoresist IP-n162 has very high refractive indices above 1.6 and strong dispersion (indicated by the low Abbe number of 24.57), while IP-S has lower refractive indices around 1.5 and weaker dispersion (higher Abbe number of 50.45). We use IP-S as a widely used low dispersion photoresist in combination with IP-n162 as the 3D direct laser writing photoresist with the strongest dispersion available from NanoScribe GmbH. The big difference in the Abbe numbers is crucial for the reduction in chromatic aberrations. Abbe numbers for different photoresists are plotted in Fig. 1, indicating that many photoresists have a rather weak dispersion (Abbe number about 50) and low refractive index ($n_d$ around 1.52). IP-Dip offers a stronger dispersion and was used in the past to create 3D printed Fraunhofer achromats [23], but IP-n162 exhibits stronger dispersion and higher refractive indices. Other possible high refractive index materials have been demonstrated utilizing nanoparticles to increase the refractive indices and dispersion [25].

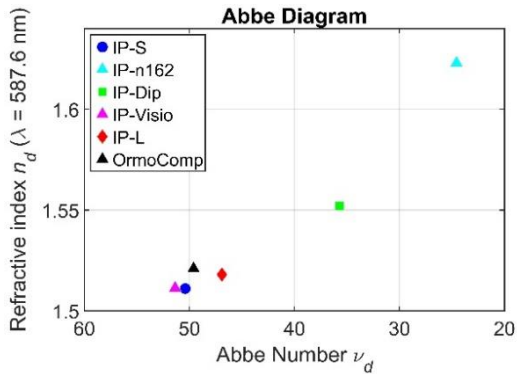

**Fig. 1.** Abbe diagram containing IP-S, IP-n162, IP-Dip, IP-Visio, IP-L, and OrmoComp. The two photoresists IP-S and IP-n162 have different refractive indices and dispersion. Due to the different Abbe numbers they can be combined to realize achromatic lenses.

In the following we discuss the designs of the different lenses as well as the fabrication results. We demonstrate the achromatic function of a refractive surface combined with a diffractive surface, as well as the combination of both approaches, which result in the first 3D direct laser written refractive/diffractive multi-material apochromat, reducing the chromatic aberrations even further. For reference we plot the LCA that is usually present in 3D printed micro-optics. We use for this reference purpose a simple aspheric design without any chromatic correction.

Fig. 2 depicts sketches of the beam paths for all three lenses as well as microscope images of the fabricated structures. The aspheric lens in fig. 2 (a) exhibits a smooth surface while the steps of the diffractive surface are clearly visible in the refractive/diffractive achromat in fig. 2(b). We use the photoresist IP-n162 for the asphere, the achromat, and for the base of the apochromat, while the top is made of IP-S. The step height is 1.1749 μm for the apochromat (IP-S surface) and 0.897 μm for the achromat (IP-n162 surface). We fabricate the microlenses using a Nanoscribe Photonic Professional GT (NanoScribe GmbH) direct laser writing machine on indium tin oxide (ITO) coated glass substrates. The lenses all have a diameter of 400 μm and a height of about 100 μm. The designed back focal length of all lenses is $f = 1$ mm to enable comparison of their optical performance. The design wavelengths for the achromatization are 500 nm, 600 nm, and 700 nm. The optical designs were optimized using Zemax, with each surface being aspheric. At the size of our lenses, diffraction at the lens aperture tends to decrease the actual focal length. Therefore, we used a custom Matlab script for an additional wave-optical fine tuning of the focusing power of the last lens surface to compensate this effect. To prevent a converging beam going through the substrate glass inducing additional (chromatic) aberrations, we designed the lenses with the flat surface facing the collimated illumination beam. This is possible due to aspheric surfaces. The fabrication parameters are slicing 0.2 μm, hatching 0.5 μm, and scan speed 50 mm/s. We use a 25x Zeiss objective (LCI "Plan-Neofluar" 25x/0.8). The used laser power for IP-n162 is 35 % (average intensity of 100 % is about 58 mW, with a repetition rate of 80 MHz the pulse energy is 254 pJ) while we use a laser power of 20 % (pulse energy 145 pJ) for the top of the apochromat of IP-S. After printing the bottom part, the sample is developed in mr-Dev 600 (20 min) to remove residual photoresist and cleaned in Isopropanol (5 min). For alignment, the structure has a small hole in the middle. The top part is printed over the created bottom part and the sample is developed again. Exposure settings are optimized by writing test arrays with different combinations of laser power and scan speed. The laser power for IP-n162 needs to be tested carefully, as the polymerization intensity window is comparably small. Too high laser intensities can lead to micro-explosions during the writing process. The same needs to be considered for IP-S, where we usually use higher laser powers but during the multi-component writing the top lens is written over already polymerized photoresist and the laser power needs to be set lower to avoid over-exposure.

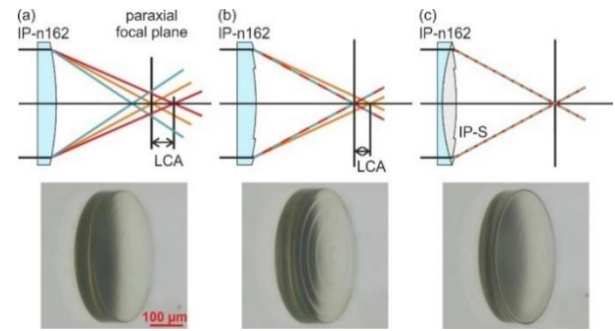

**Fig. 2.** On the left: Ray sketch for (a) aspheric lens consisting of IP-n162, (b) refractive/diffractive achromat consisting of IP-n162, and (c) refractive/diffractive apochromat consisting of IP-n162 (base) and IP-S (top). The different focal lengths for different wavelengths due to dispersion are present as LCA in the aspheric ray path. Combining refractive and diffractive surfaces reduces the LCA, while the secondary spectrum is still present. The apochromat combines refractive/diffractive surfaces as well as two different photoresists with different dispersion to correct this secondary spectrum. On the right: Microscope images of the fabricated lenses. The aspheric lens is quite smooth (a) while the steps of the diffractive surface are clearly visible the refractive/diffractive achromat (b) and on the edge of the apochromat (c).

Lower laser power also reduces possible detachment between the base and top of the lens. To compensate for the lower degree of polymerization due to the low laser power in the top part, we use additional UV exposure after the writing process (10 min UV exposure with a DymaxBlueWave 50 delivering 365 nm with an intensity of 3000 mW/cm2, at a distance of 3 cm with resulting intensity of 250 mW/cm2) and post baking at 100 °C for one h. The piezo settling time that defines the waiting time between written layers was set to five seconds except for the bottom part of the apochromat, where it was set to one second. This parameter also has an impact on the resulting optical surfaces. Too short waiting times can result in a proximity effect (depending on the writing time and geometry of the layers) thus changing the optical surface inhomogeneously while too long waiting times reduce the effective writing power.

In the following, we demonstrate the achromatization possibilities for the different 3D printed micro-optics by measuring their optical performance. For the measurements we use a collimated white light laser source (NKT SuperK-Extreme) with a dedicated monochromator (Select+) illuminating the microlens. The resulting intensity distribution is imaged with an optical microscope (Nikon TE2000-U) using a 60x, NA 0.7 objective onto a CCD camera (Allied Vision GC2450c).

the microlenses and image the resulting spot profiles by scanning the microscope objective through the focus using a PIFOC piezo nanofocusing system (Physik Instrumente).

To obtain the intensity distribution along the optical axis we perform the measurement for each wavelength +/- 55 μm around the designed paraxial focal length. The PIFOC has a step width of about 0.12 μm with a small nonlinearity which was compensated by a calibration measurement. Fig. 3 displays the measurement results of the optical performance. The strong LCA in the aspheric lens is clearly visible separating the spot location along the propagation axis, where the focal spot for 500 nm is 28 μm before the spot for 600 nm and the spot for 700 nm is 18 μm behind it. The refractive/diffractive achromat shows almost the same spot location for 500 nm and 700 nm along the propagation axis, while the secondary spectrum is visible for 600 nm with a displacement of about 5 μm. The apochromatic lens offers the smallest LCA with below 1 μm reducing the LCA even further. Our results show good agreement when compared with wave optical simulations of the LCA.

Figure 4 depicts a 2D representation of the through focus measurements for all lenses and wavelengths. The strong displacement of the focal spot along the propagation axis is nicely visible for the aspheric lens.

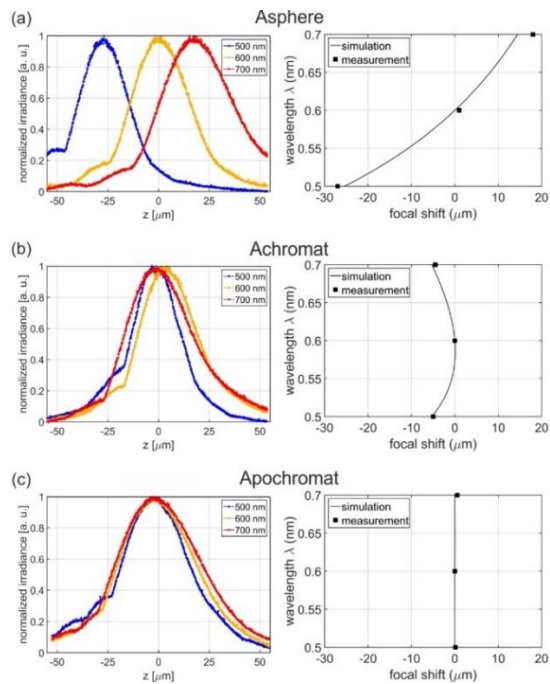

**Fig. 3.** Through focus measurement on the left for (a) the aspheric lens, (b) the refractive/diffractive achromat, and (c) the refractive/diffractive multi-material apochromat at 500 nm, 600 nm, and 700 nm. The intensity distributions along the propagation axis are shown for the different wavelength in the corresponding colors. On the right: comparison of the simulated and measured LCA of the different microlenses. The strong LCA over 46 μm along the propagation axis is clearly visible in the aspheric lens. The achromat exhibits almost the same spot location for 500 nm and 700 nm, while the secondary spectrum is still visible for 600 nm with about 5 μm z-mismatch. The refractive/diffractive multi-material apochromat offers the smallest LCA below 1 μm.

To obtain the LCA we measure the position of maximum intensity for 500 nm, 600 nm, and 700 nm. To this end, we propagate the collimated monochromatic laser beam at each wavelength through

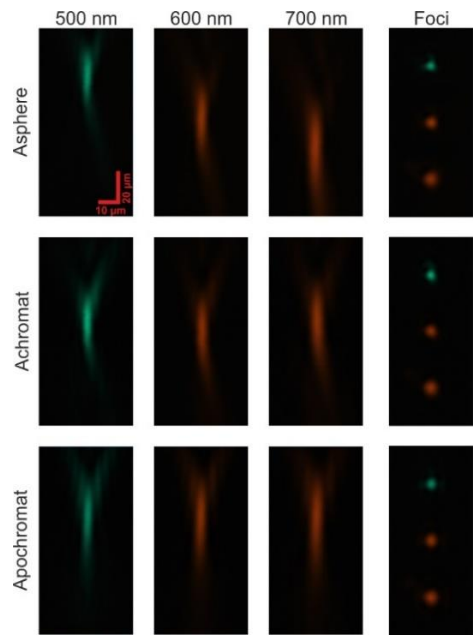

**Fig.4.** 2D through focus measurements for the different lenses at 500 nm, 600 nm, and 700 nm. The strong displacement of the focal spot along the propagation axis is clearly visible for the aspheric lens. For the refractive/diffractive achromat the spots are much closer together, however the spot for 600 nm shows the expected secondary spectral displacement compared to the other two spots, while the spots are aligned best for the apochromat. The focal spots for each lens and wavelength are depicted on the right.

For the refractive/diffractive achromat the spots are much closer together, however the spot for 600 nm shows the expected secondary spectrum displacement compared to the other two spots while the spots are aligned the best for the apochromat. The focal images are also displayed in figure 4 with a diameter of about 5 μm for 500 nm, 6 μm for 600 nm and 7 μm for 700 nm.

Figure 5 depicts imaging results of an USAF test target for all lenses. As expected the aspheric singlet displays clearly visible red/blue color seams. The achromat shows less color seams, while they do not vanish completely. The best color accuracy is offered by the apochromat, where the color seams are not visible anymore. However, monochromatic field-dependent aberrations are present, reducing the sharpness of the images. This is due to the fact that the NA 0.22 lens design was optimized for on-axis performance and did not offer field aberration corrections. This can be achieved in the future with an additional field lens whose additional chromatic aberrations would be compensated by the multi-material doublet. The blurrier on-axis picture for the apochromat compared to the achromat is probably due to the more complex design with several factors possibly influencing the image negatively (additional photoresist and interface between the resists, slight delamination between the two materials).

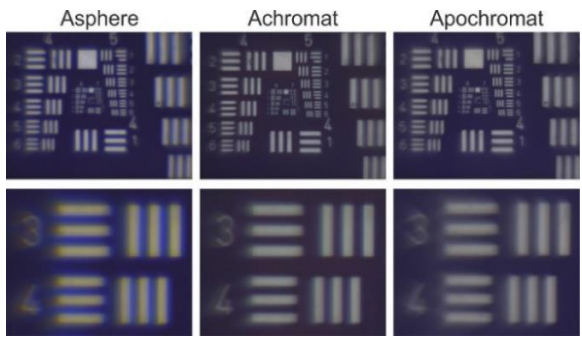

**Fig. 5.** Imaging of an USAF test target for the three different lenses. The asphere shows significant transverse chromatic aberration visible as red/blue color seams in the image. The achromat offers much smaller transverse chromatic aberration, however small color seams are still visible. In the imaging of the apochromat color seams are not visible anymore.

In summary, we demonstrated the correction of chromatic aberrations in 3D direct laser written micro-optics. Therefore, we manufactured a 3D direct laser written refractive/diffractive achromat, reducing chromatic aberration by utilizing the strong negative dispersion of the diffractive surface. This approach is very fast and enables a simple integration of achromatic function in 3D direct laser written optics paving the way for the realization of a variety of micro-optical systems using the combination of refractive and diffractive surfaces.

We further combined the refractive/diffractive approach with the combination of different photoresists (IP-n162 and IP-S) to create the first 3D direct laser written refractive/diffractive multi-material apochromat. This design reduces the LCA even further and overcomes the secondary spectrum of the apochromat, resulting in a LCA splitting below 1 μm.

**Funding.** Deutsche Forschungsgemeinschaft (GRK 2642); Bundesministerium für Bildung und Forschung (13N10146, PRINTFUNCTION, PRINTOPTICS); Baden-Württemberg Stiftung (OPTERIAL); European Research Council (3DPRINTEDOPTICS, COMPLEXPLAS).

**Acknowledgement.** We thank Alexander Quick from Nanoscribe GmbH for providing us with the novel photoresist IP-n162.

**Disclosures.** The authors declare no conflicts of interest.

**Data Availability.** Data underlying the results presented in this paper are not publicly available at this time but may be obtained from the authors upon reasonable request.

**References**

1. M. Malinauskas, M. Farsari, A. Piskarskas, and S. Juodkazis, Phys. Rep. **533**, 1 (2013).
2. M. Malinauskas, A. Zukauskas, V. Purlys, K. Belazaras, A. Momot, D. Paipulas, R. Gadonas, A. Piskarskas, H. Gilbergs, I. Sakellari, M. Farsari, and S. Juodkazis, J. Opt **12**, 8 (2010).
3. P.-I. Dietrich, M. Blaicher, I. Reuter, M. Billah, T. Hoose, A. Hofmann, C. Caer, R. Dangel, B. Offrein, U. Troppenz, M. Moehrle, W. Freude, and C. Koos, Nat. Photonics **12**, 241 (2018).
4. M. Deubel, G. von Freymann, M. Wegener, S. Pereira, K. Busch, and C. M. Soukoulis, Nat. Mater. **3**, 444 (2004).
5. A. Landowski, D. Zepp, S. Wingerter, G. Von Freymann, and A. Widera, APL Photonics **2**, 106102 (2017).
6. S. Lightman, G. Hurvitz, R. Gvishi, and A. Arie, Optica **4**, 605 (2017).
7. B. Chen, D. Claus, D. Russ, and M. R. Nizami, Opt. Lett. **45**, 5583 (2020).
8. S. Thiele, T. Gissibl, H. Giessen, and A. M. Herkommer, Opt. Lett. **41**, 3029 (2016).
9. A. Bertoncini, S. P. Laptenok, L. Genchi, V. P. Rajamanickam, and C. Liberale, J. Biophotonics e202000219 (2020).
10. C. Liberale, P. Minzioni, F. Bragheri, F. De Angelis, E. Di Fabrizio, and I. Cristiani, Nat. Photonics **1**, 723 (2007).
11. A. Asadollahbaik, S. Thiele, K. Weber, A. Kumar, J. Drozella, F. Sterl, A. M. Herkommer, H. Giessen, and J. Fick, ACS Photonics (2019).
12. L. Bremer, K. Weber, S. Fischbach, S. Thiele, M. Schmidt, A. Kaganskiy, S. Rodt, A. Herkommer, M. Sartison, S. L. Portalupi, P. Michler, H. Giessen, and S. Reitzenstein, APL Photonics **5**, (2020).
13. P. Dietrich, G. Göring, M. Trappen, M. Blaicher, W. Freude, T. Schimmel, H. Hölscher, and C. Koos, Small **16**, 1904695 (2020).
14. V. Melissinaki, M. Farsari, and S. Pissadakis, Fibers **5**, 1 (2017).
15. S. Thiele, K. Arzenbacher, T. Gissibl, H. Giessen, and A. M. Herkommer, Sci. Adv. **3**, e1602655 (2017).
16. S. Ristok, S. Thiele, A. Toulouse, A. M. Herkommer, and H. Giessen, Opt. Mater. Express **10**, 2370 (2020).
17. J. Li, S. Thiele, B. C. Quirk, R. W. Kirk, J. W. Verjans, E. Akers, C. A. Bursill, S. J. Nicholls, A. M. Herkommer, H. Giessen, and R. A. McLaughlin, Light Sci. Appl. **9**, 124 (2020).
18. J. Li, P. Fejes, D. Lorenser, B. C. Quirk, P. B. Noble, R. W. Kirk, A. Orth, F. M. Wood, B. C. Gibson, D. D. Sampson, and R. A. McLaughlin, Sci. Rep. **8**, 14789 (2018).
19. T. Gissibl, S. Thiele, A. Herkommer, and H. Giessen, Nat. Photonics **10**, 554 (2016).
20. S. Thiele, C. Pruss, A. M. Herkommer, and H. Giessen, Opt. Express **27**, 35621 (2019).
21. J. Fraunhofer, Ann. Phys. **56**, 264 (1817).
22. D. C. O'Shea, T. J. Suleski, A. D. Kathman, and D. W. Prather, (SPIE, 2003).
23. M. Schmid, S. Thiele, A. Herkommer, and H. Giessen, Opt. Lett. **43**, 5837 (2018).
24. M. Schmid, D. Ludescher, and H. Giessen, Opt. Mater. Express **9**, 4564 (2019).
25. K. Weber, D. Werdehausen, P. König, S. Thiele, M. Schmid, M. Decker, P. W. De Oliveira, A. Herkommer, and H. Giessen, Opt. Mater. Express **10**, 2345 (2020).